\documentstyle[12pt,aps]{revtex}%
\begin{document}

\newcommand{\bce}{\begin{center}}
\newcommand{\ece}{\end{center}}
\newcommand{\be}{\begin{equation}}
\newcommand{\ee}{\end{equation}}
\newcommand{\bea}{\vspace{0.25cm}\begin{eqnarray}}
\newcommand{\eea}{\end{eqnarray}}
\def\NCA{{Nuovo Cimento } A }
\def\NIM{{Nucl. Instrum. Methods}}
\def\NPA{{Nucl. Phys.} A }
\def\PLA{{Phys. Lett.}  A }
\def\PRL{{Phys. Rev. Lett.} }
\def\PRA{{Phys. Rev.} A }
\def\PRC{{Phys. Rev.} C }
\def\PRD{{Phys. Rev.} D }
\def\ZPC{{Z. Phys.} C }
\def\ZPA{{Z. Phys.} A }
\def\PTP{{Progr. Th. Phys. }}
\def\LNC{{Lett. al Nuovo Cimento} }

{\bf Experimental test of local realism using
 non-maximally entangled states }
\vskip 0.5cm
{Marco Genovese, G. Brida and C.Novero}
\vskip 0.2cm
{\it Istituto Elettrotecnico Nazionale Galileo Ferraris, 
Str. delle Cacce 91,  I-10135 Torino}

\vskip 0.5cm

{E. Predazzi}
\vskip 0.2cm
{ \it Dip. Fisica Teorica Univ. Torino and INFN, via P. Giuria 1,  I-
10125 Torino }

\vskip 0.2cm
{\bf Abstract}

{ In this paper we describe a test of Bell inequalities using a non-
maximally entangled state, which represents an important step in the direction of  eliminating  the detection loophole.
The experiment is based on the creation of a polarisation entangled state via 
the superposition, by use of an appropriate optics, of the spontaneous 
fluorescence emitted by two non-linear crystals driven by the same pumping 
laser.  }

\vskip 1cm
Entanglement is the main resource of Quantum Information Processing and thus deserves to be widely studied. One of the most characteristic properties of entanglement are the correlations between space-like separated entangled systems   studied by Einstein-Podolsky-Rosen, for which Bell showed that no Local Hidden Variable Theory can reproduce all the results of Quantum Mechanics and, in particular, that some inequality, valid for every LHV theory, is violated by QM.

Many interesting experiments have been devoted to a test of Bell 
inequalities, the most interesting of them using photon pairs 
\cite{Mandel,asp,franson,type2}, leading to a substantial agreement with 
quantum mechanics and disfavouring LHV theories. But, up to now, no experiment 
has yet been able to exclude definitively such theories {\cite{gar}}, because of the low total detection efficiency, which requires the hypothesys that the 
observed sample of particle pairs is a faithful subsample of the initial set of pairs. 
In the 90's  a big progress in the direction of eliminating this loophole has 
been obtained by using parametric down conversion (PDC) processes, which 
overcomes some former limitations, as the poor angular 
correlation of atomic cascade photons { \cite{asp}}.
The first experiments were performed with type I PDC, 
which gives phase and momentum entanglement and can be used for a test of Bell 
inequalities using two spatially separated interferometers 
 {\cite{franson}}. The use of beam splitters, however, strongly 
reduces the total quantum efficiency. 
In alternative, a polarisation entangled state can be generated  {\cite{ou}}, 
but in most of the used configurations,  
half of the initial photon flux is lost. 

Recently, an experiment where a polarisation entangled state is directly 
generated, has been realised using Type II PDC  {\cite{type2}}. This scheme has
 permitted, at the price of delicate compensations for having identical arrival time of the ordinary and extraordinary photon, a much higher total efficiency 
than the previous ones, which is, however, still far from the value of $0.81$ 
required for eliminating the detection loophole for a maximally entangled state. A large 
interest remains therefore for new experiments increasing total quantum 
efficiency in order to reduce and finally overcome the efficiency loophole. 

For this purpose, we have considered  {\cite{napoli}} the possibility of generating 
a polarisation entangled state via the
superposition of the spontaneous fluorescence emitted by two non-linear crystals 
(rotated for having orthogonal polarisation) driven
by the same pumping laser  {\cite{hardy}}. The crystals are put in cascade
along the propagation direction of the pumping laser and the superposition
is obtained by using an appropriate optics. If the path between the two
crystal is smaller than the coherence length of the laser, the two photons
path are indistinguishable and a polarisation entangled state is created.
The possibility of easily obtaining a non maximally entangled state 
 is very important, for it has been recognised that 
in this case the lower limit on the total detection 
efficiency for eliminating the detection loophole is as small as 0.67 
 {\cite{eb}}.

%\section{DESCRIPTION OF THE EXPERIMENT}

The experimental set-up is composed of two crystals of 
$LiIO_3$ placed along the pump laser propagation, 250 mm apart,  
a distance smaller than the coherence length of the pumping laser. This 
guarantees indistinguishability in the creation of a couple of photons in the 
first or in the second crystal. A couple of planoconvex lenses of 120 mm focal 
length centred in between focalises the spontaneous emission from the first 
crystal into the second one. A hole of 4 mm diameter is drilled into the centre of the lenses to 
allow transmission of the pump. A small quartz plate  compensates the displacement of the pump due to birefringence in the first crystal.  
Finally, a half-wavelength plate immediately after the condenser rotates the 
polarisation of the Argon beam. 
We have used as photo-detectors two avalanche photodiodes with active quenching 
(EG\&G SPCM-AQ) coupled to an optical fiber.

A very interesting degree of freedom of this configuration is given by the fact 
that by tuning the pump intensity between the two crystals, one can easily
tune the ratio $f$ between the two component of the entangled state ($f=1$ for a maximally entangled state). Furthermore, every Bell state can be generated (for example for applications to quantum information processing) changing the phase between the two components (moving the second crystal) and/or changing the polarisation on one of the branches after the second crystal.  
The main difficulty of this configuration is in the alignment, which is of 
fundamental importance for having a high visibility. 
This problem has been solved using {\cite{brida}}   an 
optical amplifier
scheme, where a solid state laser is injected into the crystals together with
the pumping laser.
We think that the proposed scheme is well suited for leading to a further step 
toward a conclusive experimental test of non-locality in quantum mechanics. The
main advantage of the proposed configuration with
respect to most of the previous experimental set-ups is that all the
entangled pairs are selected (and not only $< 50 \%$ as with beams splitters), 
furthermore it does not require delicate compensations for the optical paths ofthe ordinary and extraordinary rays after the crystal.

At the moment, the results which we are going to present are still far from
a definitive solution of the detection loophole; nevertheless, being the first 
test of Bell inequalities using a non-maximally
entangled state, they
represent an important step in this direction. Furthermore, this
configuration permits to use any pair of correlated frequencies and not only
the degenerate ones. We have thus realised this test using for a first time
two different wave lengths (at $633$ and $789$ nm).

It must be acknowledged that a set-up for generating polarisation entangled 
pairs of photons, which presents analogies with our, has been realised recently 
in Ref.  {\cite{Kwiat}}
The main difference between the two experiments is that in \cite{Kwiat} the
two crystals are very thin and in contact with orthogonal optical
axes: this permits a "partial" { \cite{ser}} superposition of the two emissions with
opposite polarisation. This overlapping is mainly due to the finite dimension of 
the pump laser beam, which reflects into a finite width of each wavelength 
emission.
A much better superposition can be obtained with the present scheme.
More recently  {\cite{k2}}, they have also performed a test about local realism 
using a particular kind of Hardy equalities  {\cite{HardyE}}. Their result is in 
agreement with quantum mechanics modulo the detection loophole, no discussion 
concerning the elimination of loopholes for this equality is presented.   

As a first check of our apparatus, we have measured the interference
fringes, varying the setting of one of the polarisers, leaving the other
fixed. We have found a high visibility, $V=0.973 \pm 0.038$.

Our results are summarised by the value obtained for the Clauser-Horne sum 

\begin{equation}
CH=N(\theta _{1},\theta _{2})-N(\theta _{1},\theta _{2}^{\prime })+N(\theta
_{1}^{\prime },\theta _{2})+N(\theta _{1}^{\prime },\theta _{2}^{\prime
})-N(\theta _{1}^{\prime },\infty )-N(\infty ,\theta _{2})  \label{eq:CH}
\end{equation}
which is strictly negative for local realistic theory. In (\ref{eq:CH}), 
$N(\theta _{1},\theta _{2})$ is the number of coincidences between
channels 1 and 2 when the two polarisers are rotated to an angle $\theta _{1}$
and $\theta _{2}$ respectively ($\infty $ denotes the absence of selection
of polarisation for that channel).
On the other hand,  quantum mechanics predictions for $CH$ can be larger than 
zero.
In our case we have generated a state with $f \simeq 0.4$: in this case the 
largest violation of the inequality is reached for $\theta_1 =72^o.24$ , 
$\theta_2=45^o$, $\theta_1 ^{\prime}= 17^o.76$ and 
$\theta_2 ^{\prime}= 0^o$. Our experimental result is $CH = 512 \pm 
135$ coincidences per second, 
which is almost four standard deviations different from zero and compatible with 
the theoretical value predicted by quantum mechanics.

\end{document}